\documentclass[12pt]{article}
\usepackage{amsfonts,amsmath,amssymb,epsf,subfigure}
\usepackage{graphicx,color}
\def\CN{{\cal N}}
\def\t{\theta}
\def\Om{\Omega}

\newcommand{\de}{\partial}
\newcommand{\be}{\begin{equation}}
\newcommand{\ba}{\begin{eqnarray}}
\newcommand{\ea}{\end{eqnarray}}
\newcommand{\ee}{\end{equation}}

\newcommand{\f}{\frac}
\newcommand{\s}{\sqrt}

\newcommand{\ti}{\tilde}
\newcommand{\ap}{\alpha}

\newcommand{\ddd}{\cdot\cdot\cdot}
\newcommand{\no}{\nonumber \\}

\begin{document}

\begin{titlepage}
\thispagestyle{empty}
KUNS-2144
\begin{flushright}
\end{flushright}

\bigskip

\begin{center}
\noindent{\large \textbf{Geometric Entropy and Hagedorn/Deconfinement Transition}}\\
\vspace{2cm} \noindent{Mitsutoshi
Fujita\footnote{e-mail:mfujita@gauge.scphys.kyoto-u.ac.jp}, Tatsuma
Nishioka\footnote{e-mail:nishioka@gauge.scphys.kyoto-u.ac.jp}
 and Tadashi
Takayanagi\footnote{e-mail:takayana@gauge.scphys.kyoto-u.ac.jp}}\\
\vspace{1cm}

 {\it  Department of Physics, Kyoto University, Kyoto 606-8502, Japan
 }

\vskip 3em
\end{center}

\begin{abstract}

It has recently been proposed that the entanglement entropy can be
an order parameter of confinement/deconfinement transitions. To find
a clear evidence, we introduce a new quantity called the geometric
entropy, which is related to the entanglement entropy via a double
Wick rotation. We analyze the geometric entropy and manifestly show
that its value becomes discontinuous at the Hagedorn temperature
both in the free $\CN =4$ super Yang-Mills and in its supergravity
dual.

\end{abstract}

\end{titlepage}

\section{Introduction}
\setcounter{equation}{0} {\hspace{5mm}}
The entanglement entropy has been playing very important roles in
recent studies of quantum field theories motivated by both string
theory and condensed matter theory. The entanglement entropy $S_A$
can be regarded as a measure of degree of freedom confined in a
certain space-like region $A$, chosen arbitrarily. In the two
dimensional conformal field theories, it is indeed proportional to
the central charge \cite{HLW,CC}. In quantum field theory with UV
fixed points, we can in general show that the leading ultraviolet
divergent term of $S_A$ is proportional to the area of the boundary
of $A$ \cite{Area}, while the subleading terms depend on the shape
of the region $A$. Even though the direct computation of $S_A$ often
involves complicated analysis, the holographic calculation \cite{RT}
based on AdS/CFT correspondence \cite{Ma} provides a more tractable
way of doing this (for recent progresses see e.g.
\cite{Em}-\cite{Mi}).

In the analysis of quantum phase transitions which frequently appear
in condensed matter systems, it has been pointed out that the
entanglement entropy can be used as a quantum order parameter
\cite{Vidal,KP,LW}. Especially it is quite useful to specify the
phases in topological theories as ordinary correlation functions
become trivial \cite{KP,LW}, while the entanglement entropy (called
topological entanglement entropy\footnote{Refer to \cite{PP} for a
holographic calculation of topological entanglement entropy.}) does
not.

In the recent papers \cite{NT,KKM}, the entanglement entropy in
confining gauge theories has been studied holographically and it has
been shown that it undergoes a sort of phase transition when we
change the size of $A$. This behavior has been confirmed recently in
the lattice gauge theories \cite{Ve,BP}. Thus it is natural to
expect that the entanglement entropy can be an order parameter of
the confinement/deconfinement transition in gauge theories
\cite{NT,KKM}. In order to reinforce this idea, the main purpose of
this paper is to show that the entanglement entropy (or more
generally, the geometric entropy) is a nice order parameter of the
confinement/deconfinement transition in $\CN =4$ super Yang-Mills
theory when we change the temperature. This phase transition is
well-known to be dual to the Hagedorn transition in string theory
via the AdS/CFT.

Especially we will employ the free Yang-Mills analysis in
\cite{AMMPR} and compute a certain entropy defined later, which is
closely related to the entanglement entropy. This quantity is not
exactly the same as the ordinary entanglement entropy, but can be
regarded as its double Wick rotated one. Since it is defined
geometrically, we will call this quantity the geometric entropy in
this paper\footnote{In some literature, the authors defined the
geometric entropy to be exactly the same as the entanglement
entropy. However, in this paper, we use the term `geometric entropy'
in a broader sense.}. The relation between our geometric entropy and
the entanglement entropy is analogous to the one between the Wilson
loop and the Polyakov loop. Remember that the Polyakov loop is
strictly speaking not well-defined on a compact manifold because it
introduces a positive charge, while our geometric entropy is
well-defined. It has also a nice property that its gravity dual is
easy to analyze. We explicitly examine its behavior and show that
its value jumps at the transition point. This behavior qualitatively
agrees with the results in the supergravity dual.
 We will also discuss that this quantity can be a useful order parameter in
other theories such as the topological field theories and two
dimensional Yang-Mills theory.

This paper is organized as follows: In section two, we define the
geometric entropy as a double Wick rotation of the entanglement
entropy. We also calculate this quantity holographically in the
$AdS_5$ back hole background. In section three, we compute the
geometric entropy in the free $\CN =4$ Yang-Mills theory and compare
the results with its dual gravity result. In section four, we
briefly discuss the application of the geometric entropy to
topological field theories and two dimensional Yang-Mills. In
section five, we summarize our conclusion.

\section{Geometric Entropy in $\CN =4$ SYM and $AdS/CFT$}
\setcounter{equation}{0}
\subsection{Definition of Geometric Entropy}
We compactify on  $S^3$ a four dimensional quantum field theory such
as the $\CN =4$ super Yang-Mills. At finite temperature
$T=\f{1}{\beta}$, it is defined on $S^1\times S^3$. We express the
metric of $S^3$ as follows \be
d\Omega^2_{(3)}=d^2\theta+\sin^2\theta(d\psi^2+\sin^2\psi d\phi^2),
\label{threes} \ee where $0\le \theta,\psi \le \pi$ and $0\le \phi
\le 2\pi$.

Now we change the periodicity of $\phi$ into $0\leq \phi \leq 2\pi
k$. For $k\neq 1$, there exist conical singularities at $\psi=0,\pi$
with the deficit angle $\delta=2\pi(1-k)$. The submanifold of $S^3$
defined by these singular points is equal to $S^1$ (the largest
circle of $S^3$).

The partition function on this singular space is defined to be
$Z_{YM}(k)$. The ordinary partition function on $S^1\times S^3$
coincides with $Z_{YM}(1)$. We can consider the normalized partition
function and can regard it as follows \be
\f{Z_{YM}(k)}{(Z_{YM}(1))^{k}}=\mbox{Tr} \rho^k, \ee where
$\rho=e^{-2\pi H}$ is the density matrix when we regard the
coordinate $\phi$ as the Euclidean time (and $H$ is its Hamiltonian)
via the double Wick rotation.

Then, following the definition of von-Neumann entropy, we define the
geometric entropy $S_{G}$ by \be S_{G}=-\mbox{Tr}\rho\log \rho=
-\f{\de}{\de k}\log\left[\f{Z_{YM}(k)}{(Z_{YM}(1))^{k}}\right]
\Biggr|_{k=1}.\label{sdef}\ee As is clear from the above, the
geometric entropy is different from the entanglement entropy but is
related to it via the double Wick rotation. In other words, in the
ordinary entanglement entropy we regard the thermal circle $S^1$ is
the Euclidean time, while in our geometric entropy we regarded
$\phi$ in $S^3$ as the Euclidean time. Thus it is analogous to the
Polyakov loop instead of the Wilson loop. Remember that the Polyakov
loop is strictly speaking not well-defined on a compact manifold,
while our geometric entropy is well-defined. It will also be an
interesting future problem to compute the ordinary entanglement
entropy to see if it can be an order parameter, though this will
require a more complicated analysis.

In the practical computations of $S_G$, it is convenient to take the
values of $k$ to be fractional $k=\f{1}{n}$. This theory is
equivalent to the $\CN =4$ super Yang-Mills on the orbifold $S^3/Z_n$.
The $Z_n$ identification is simply defined by
$\phi\sim\phi+\f{2\pi}{n}$ in the coordinate (\ref{threes}). This is
locally the same as the $\mathbb{C}/Z_{n}$ orbifold and thus is
non-supersymmetric. The careful analysis of the spin structure
\cite{Adams:2001sv} shows that $n$ should be an odd integer. Then we
can calculate the geometric entropy from the formula \be
 S_{G}=
-\f{\de}{\de(1/n)}\log\left[\f{Z_{YM}(S^3/Z_n)}{(Z_{YM}(S^3))^{1/n}}\right]
\Biggr|_{n=1}. \label{sdeff}\ee

The partition function in the $\CN =4$ Yang-Mills theory at finite
temperature and coupling has not been obtained so far. Therefore,
in the next section, we will perform the calculation of the
partition function and find the geometric entropy in the free $\CN =4$
Yang-Mills theory. As we will see later, even under this free field
approximation, we can still reproduce qualitative behavior of $S_G$
expected from the gravity computation.

\subsection{Holographic Calculation of $S_G$}
We would like to first compute $S_G$ in the dual gravity side. In
the Yang-Mills language, the supergravity analysis is dual to the
strongly coupled Yang-Mills.

If we require that the boundary is given by $S^1\times S^3$, then
only two examples are known as the bulk space \cite{Witten}; one is
the thermal AdS \be ds^2=\left(\f{r^2}{R^2}+1\right)d\tau^2
+\f{dr^2}{\f{r^2}{R^2}+1}+r^2d\Omega^2_{(3)}, \label{thermalads} \ee
and the other one is the AdS (large) black hole \be
ds^2=\left(\f{r^2}{R^2}+1
-\f{M}{r^2}\right)d\tau^2+\f{dr^2}{\f{r^2}{R^2}+1-\f{M}{r^2}}
+r^2d\Omega^2_{(3)}. \label{adsbh} \ee The horizon of the latter
spacetime is at $r_+$ defined by $\f{r_+^2}{R^2}+1 -\f{M}{r_+^2}=0$.

By requiring the smoothness of the Euclidean geometry, we find that
the periodicity of $\tau$ (\ref{adsbh}) is given by \be\label{Temp}
\ti{\beta}=\f{2\pi r_+ R^2}{2r_+^2+R^2}. \ee The analysis of the
free energy shows that at low temperature $\ti{\beta}>\ti{\beta_H}$
the thermal AdS solution is stable, while at high temperature
$\ti{\beta}<\ti{\beta_H}$ the AdS black hole solution becomes
favored \cite{Witten}. Here the phase transition temperature is
given by $\beta_H=\f{2\pi}{3}R$ and is known as the Hawking-Page
transition.

Now we would like to compute $S_G$. In order for this we need to put
the deficit angle $2\pi(1-\ap)$ along the circle $S^1$ on $S^3$. The
presence of the codimension two deficit angle leads to the delta
functional source of the scalar curvature $R=4\pi(1-\ap)\delta(x)$.
If we plug this into the Einstein-Hilbert action, we get \be
S_{sugra}=-\f{1}{16\pi G^{(5)}_N}\int
\s{g}R+\ddd=-\f{\mbox{Area}(\gamma)}{4G^{(5)}_N}(1-\ap), \ee where
$\gamma$ is the codimension two surface where the deficit angle is
localized (these arguments are very similar to the one in
\cite{Fu}). Using the bulk to boundary relation
$Z_{CFT}=Z_{sugra}=e^{-S_{sugra}}$ in the supergravity approximation
\cite{Gubser:1998bc, Witten:1998qj}, we eventually be able to obtain
the geometric entropy as follows \be S_G=-\f{\de}{\de\ap}\log
\f{Z_{sugra}(\ap)}{(Z_{sugra}(0))^\ap}= \f{\de}{\de\ap}S_{sugra}
=\f{\mbox{Area}(\gamma)}{4G^{(5)}_N}. \ee

The surface $\gamma$ for the geometric entropy $S_G$ is given by the
codimension three surface defined by $\sin\psi=0$, which extends in
the $(\tau,r,\theta)$ direction. We put the UV cut off at
$r=r_{\infty}\gg R$.

In the thermal AdS, we find \be
S^{ads}_G=\f{1}{4G^{(5)}_N}\int^{\ti{\beta}}_0 \int^{r_{\infty}}_0
rdr\int^{2\pi}_{0} d\theta=\f{\pi\ti{\beta}
r^2_{\infty}}{4G^{(5)}_N}, \ee while in the AdS black hole we get
\be S^{bh}_G=\f{1}{4G^{(5)}_N}\int^{\ti{\beta}}_0
\int^{r_{\infty}}_{r_+} rdr\int^{2\pi}_{0} d\theta=\f{\pi\ti{\beta}
(r^2_{\infty}-r^2_{+})}{4G^{(5)}_N}. \ee  For the large AdS black
hole we have the relation equivalent to (\ref{Temp})\be r_+=\f{\pi R^2}{2\ti{\beta}}+
\s{\f{\pi^2R^4}{4\ti{\beta}^2}-\f{R^2}{2}}. \ee Notice that the AdS
 BH exists when $\ti{\beta}<\f{\pi R}{\s{2}}$. Below we introduce the
dimensionless temperature $\beta$ defined by \be
\beta=\f{\ti{\beta}}{R}. \ee

We are interested in the difference\footnote{In the original
arguments of the Hawking-Page transition, we needed to choose
slightly different temperature between thermal AdS and AdS BH. This
subtlety is not important in our argument.} of these entropies. This
quantity is vanishing at the temperature lower than the Hagedorn
transition, i.e. $\beta>\f{2\pi}{3}$. On the other hand, at high
temperature ($\beta<\f{2\pi}{3}$), we obtain the non-vanishing
result \be\label{GRGE} \Delta S_{G}=-\f{\pi\ti{\beta} r^2_{+}}{4G^{(5)}_N}
=-\f{\pi R^3}{4G^{(5)}_N}\cdot
\f{\left(\pi+\s{\pi^2-2\beta^2}\right)^2}{16\beta}. \ee

If we assume the $AdS_5\times S^5$ background in type IIB string
dual to the $\CN =4$ super Yang-Mills, we can rewrite the expression as
follows\footnote{In the final expression we performed the high
temperature expansion. Notice that the leading term $\sim
\beta^{-1}$ agrees with the result $\Delta S_A=-\f{\pi N^2 L}{4V_1}$
in \cite{NT} by identifying $\beta=2\pi \f{L}{V_1}$, which is
obtained by looking at conformally invariant quantities. This
agreement is because the high temperature limit $\beta\to 0$ means
that the size of the sphere $S^3$ becomes infinitely large and so we
can regard it as $R^3$.} (using $\f{R^3}{G^{(5)}_N}=\f{2N^2}{\pi}$)
\ba \Delta S_{G}&=&-\f{\pi^2
N^2}{8\beta}\left(1+\s{1-\f{2\beta^2}{\pi^2}}\right)^2 \no
 &\simeq&
 -\f{N^2}{2}\left(\f{\pi^2}{\beta}-\beta\right)+O(\beta^3).\label{sugras}
\ea

If we start with low temperature and increase the temperature
gradually, then at $\beta=\f{2\pi}{3}$, the quantity $\Delta S_{G}$
suddenly jumps from zero to $-\f{2\pi^2 N^2}{9\beta}$ (see the Figure
\ref{fig:GEGR} in the next section). Thus we can conclude that $S_G$
is an order parameter of the confinement/deconfinement transition.
Our analysis can be comparable to that of the holographic
entanglement entropy in a confining gauge theories at zero
temperature in \cite{NT,KKM,PP}. In the latter case, the derivative
of the entropy (not the entropy itself) jumps when we change the
size of the subsystem $A$ which defines the entanglement entropy
$A$.

\section{Geometric Entropy in Free $\CN =4$ super Yang-Mills}
\setcounter{equation}{0} {\hspace{5mm}}
We would like to return to the Yang-Mills analysis. Since the
evaluation of the partition function in the interacting $\CN =4$ super
Yang-Mills is rather difficult, here we would like to be satisfied
with the free Yang-Mills calculation. As noticed in \cite{AMMPR},
the free Yang-Mills analysis can capture the confinement/deconfinement
transition since the Gauss law constraint on $S^3$ restricts the
total charge to be vanishing.

The partition function in the free Yang-Mills theory is written in
the form \cite{AMMPR} (we set $x=e^{-\beta}$) \ba Z_{YM}&=&\int
[dU]~ e^{\sum_{m=1}^\infty\f{1}{m}
\left(z_{s}(x^m)+z_{v}(x^m)+(-1)^{m+1}z_{f}(x^m)\right)
\mbox{tr}(U^m)\mbox{tr}((U^{\dagger})^m) },\\  &=& \int
\prod_{i=1}^N d\theta_i e^{-\sum_{i\neq j}V(\theta_i-\theta_j)},
 \ea where we diagonalized the unitary matrix $U$ in the final
 expression. Also $z_s(x)$, $z_v(x)$ and $z_f(x)$ denote the single
 particle partition functions of scalars, vectors and fermions in a
 given gauge theory.

The potential $V(\theta)$ is given as follows \be V(\theta)=\log
2+\sum_{m=1}^\infty V_{m}\cos(m\theta), \ee where we set \be
V_{m}=\f{1}{m}(1-z_{s}(x^m)-z_{v}(x^m)-(-1)^{m+1}z_{f}(x^m)). \ee In
the $\CN =4$ SYM on the orbifold $S^3/Z_n$, assuming $n$ is an odd
integer, we find \be\label{SPF}
z_s(x)=6\f{x(1+x^n)}{(1-x)^2(1-x^n)},\ \ \ \
z_v(x)=\f{2x^2(1+2x^{n-1}-x^n)}{(1-x)^2(1-x^n)},\ \ \ \
z_{f}(x)=\f{16x^{\f{n}{2}+1}}{(1-x)^2(1-x^n)}. \ee For the detailed
derivation of these functions, please refer to the appendix A. A
calculation of $z(x)$ in a different orbifold has been done in
\cite{Hi}.

To solve the matrix model, we introduce the density of the
eigenvalues as usual \be \rho(\theta)=\f{1}{2\pi}+\sum_{m=1}^\infty
\f{\rho_{m}}{\pi}\cos(m\theta), \ee so that it is normalized as
$\int^\pi_{-\pi}d\theta \rho(\theta)=1.$  Then the free energy looks
like \be\label{FreeEn} \beta F=\beta E_{0}+N^2\sum_{m=1}^\infty
|\rho_{n}|^2V_{m}, \ee
where 
$E_0$ is the Casimir energy. The
contribution from the Casimir energy is not important for our
purpose. This is because we always subtract the result from the one
with the periodic boundary condition in the Euclidean time direction
as we did so in the gravity side and therefore the Casimir energy
part cancels out.

At low temperature (i.e.\ confining phase), we find $\rho_{n\geq
1}=0$. When $T=T_{H}$ (the Hagedorn transition point, i.e.
$V_1(x)=0$), $\rho_{1}$ jumps to $\rho_{1}=1$. On the other hand, in
the high temperature limit, the eigenvalue distribution becomes
delta-functional
$\rho(\theta)=\delta(\theta)=\f{1}{2\pi}\sum_{n=-\infty}^\infty
e^{in\theta}$ and thus $\rho_{n\geq 1}=1$.

\subsection{Analysis Near the Transition}
In the low temperature case, we can assume only
$\rho_1$ becomes non-zero.
The density of the eigenvalues can be solved as \cite{AMMPR}
\begin{align}
\rho(\theta) &= \f{1}{\pi\sin^2\left(\f{\theta_0}{2}\right)}
\s{\sin^2\left(\f{\theta_0}{2}\right) - \sin^2\left(\f\theta
    2\right)} \cos\f{\theta}{2},\label{ApDenFunc}\\
&\sin^2\left(\f{\theta_0}{2}\right) = 1- \s{1-\f{1}{z_s(x)+z_v(x)+z_f(x)}},\label{theta0}
\end{align}
where $\rho(\theta)$ has the support on $\{-\theta_0 < \theta < \theta_0\}$.

Putting (\ref{ApDenFunc}) into (\ref{FreeEn}), we find the
free energy in fairly simple form
\begin{align}\label{YMFree}
  \beta F = -N^2\left( \f{1}{2\sin^2\left(\f{\theta_0}{2}\right)} +
    \f12\log\left(\sin^2\left(\f{\theta_0}{2}\right) \right)-\f12 \right).
\end{align}
For $T>T_H$, this action is well-defined (see
(\ref{theta0})) and takes values of order $N^2$, while this is order one
for $T<T_H$ because the coefficients $V_n$ appeared in
(\ref{FreeEn}) are positive and the minimal configuration $\rho_{n\ge 1} = 0$
gives  $F = 0$.

\begin{figure}
  \centering
  \subfigure[Yang-Mills]{
    \includegraphics[height=4cm]{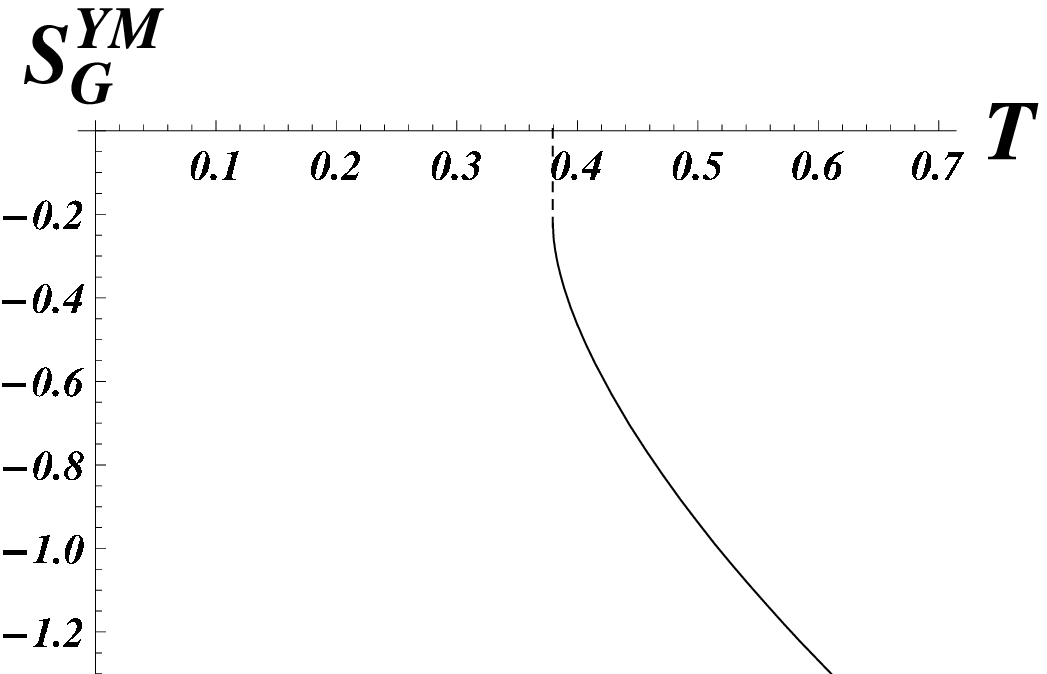}\label{fig:GEYM}
  }
  \centering
  \subfigure[Gravity]{
    \includegraphics[height=4cm]{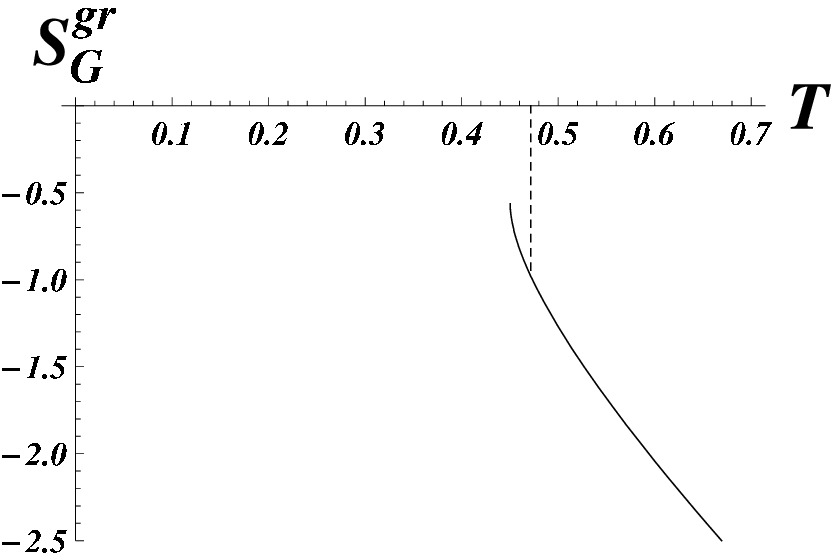}\label{fig:GEGR}
  }
  \caption{The behavior of  $\Delta S_G/N^2$ for free Yang-Mills
    (a) and IIB supergravity (b). The horizontal axis corresponds
    to the temperature $T$.
    The temperature at which the phase transition occurs is
    $T_H = -1/\ln (7-4\s3)= 0.379$ in the Yang-Mills theory and
     $T_H = 3 /2\pi = 0.477$ (dashed line) in the dual gravity.
    Notice that the line starts at $T = \s2 /\pi = 0.450$ in (b)
    above which the AdS black hole exists.
  } \
\end{figure}

The geometric entropy can be computed from the partition function
$Z$ in the free Yang-Mills by the following formula as defined in
(\ref{sdeff}) \ba\label{YMGE} \Delta
S_{G}&=&-\f{\de}{\de(1/n)}\left(\log Z(n)-\f{1}{n}\log
Z(1)\right)\Big|_{n=1}\no &=& -\f{\de}{\de n}\left( (\beta
F)(n)-\f{1}{n}(\beta F)(1)\right)\Big|_{n=1}. \ea Using the single
particle partition function (\ref{SPF}), we can easily plot the
geometric entropy near the phase transition as in Figure
\ref{fig:GEYM}, where we also plot the gravity result (\ref{sugras})
in Figure \ref{fig:GEGR}. They look qualitatively similar and show a
jump at the transition point. One may notice that the value of
$\f{dS_G}{dT}$ is infinite in the free $\CN =4$ Yang-Mills, while it
is finite in the gravity. This difference comes from the following
fact. In gravity side, the temperature at which the black hole
solution appears and the temperature at which the black hole becomes
stable against the thermal AdS, are different. However, in the free
Yang-Mills limit, they do degenerate as is clear from the fact that
there is no saddle point or local minima in the free energy
(\ref{FreeEn}).

\begin{figure}
\begin{center}
\includegraphics[height=6cm]{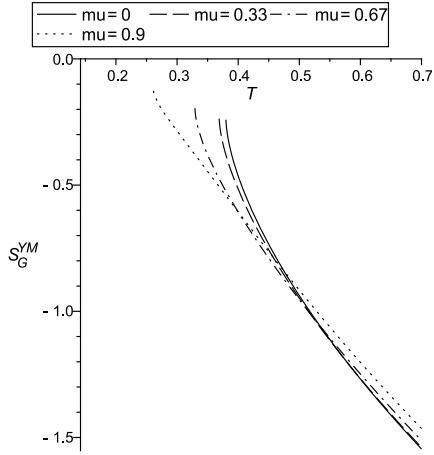}
\end{center}
\caption{The behavior of  $\Delta S_G/N^2$ for free Yang-Mills with
 the specified values of the chemical potential
 $(\mu_1,\mu_2,\mu_3)=(\mu,0,0)$ of the $R$-charges.}
\label{fig:GE}
\end{figure}

It is also straightforward to take the chemical potential
$(\mu_1,\mu_2,\mu_3)$ of the $R$-charges $(Q_1,Q_2,Q_3)$ into account
by multiplying $z_{s}(x)$ and $z_{f}(x)$ by the factor
$(e^{\mu_1}+e^{-\mu_1}+e^{\mu_2}+e^{-\mu_2}+e^{\mu_3}+e^{-\mu_3})/6$ and
$(e^{\f{\mu_1}{2}}+e^{-\f{\mu_1}{2}})(e^{\f{\mu_2}{2}}+e^{-\f{\mu_2}{2}})
 (e^{\f{\mu_3}{2}}+e^{-\f{\mu_3}{2}})/8$, respectively. The matrix
model description and its phase structure in the $R$-charged case
have been worked out in \cite{BaWa,YY,HO}. We focus on the specific
case $(\mu_1,\mu_2,\mu_3)=(\mu,0,0)$, and the result is plotted in
Figure \ref{fig:GE}. Even though the transition temperature
decreases as $\mu$ becomes large ($\mu<1$), the discontinuity of
$S_G$ is still present. A more non-trivial extension will be to
introduce a potential with respect to the rotation in the $S^3$,
which is dual to a rotating black hole. The analysis of the phase
structure in the rotating system has been recently done 
\cite{Nishioka} (the same system in decopling limit was studied in
\cite{HO2}).

\subsection{Analysis of High Temperature Limit}
In the high temperature limit $x\to 1$ ($\beta\to 0$), the
$z_{s}(x^m)$, $z_{v}(x^m)$ and $z_{f}(x^m)$ behave
like \ba && z_{s}(x^m)\simeq \f{12}{m^3
n\beta}+\f{n^2-1}{mn\beta}+O(\beta),\no && z_{v}(x^m)\simeq
\f{4}{m^3 n\beta^3}+\f{n^2-6n-1}{3mn\beta}+O(1), \no &&
z_{f}(x^m)\simeq \f{16}{m^3
n\beta^3}-\f{2(2+n^2)}{3mn\beta}+O(\beta). \ea Then we can evaluate
the free energy as follows \ba  &&-\beta F_{scalar}\simeq \f{N^2}{30
n\beta^3}(4\pi^4-5\pi^2\beta^2+5n^2\pi^2\beta^2),\no && -\beta
F_{vector}\simeq \f{N^2}{90 n\beta^3}(4\pi^4-5\pi^2\beta^2-30 n
\pi^2\beta^2+5 n^2\pi^2\beta^2),\no && -\beta F_{fermion}\simeq
\f{N^2}{90 n\beta^3}(14\pi^4-10\pi^2\beta^2-5n^2\pi^2\beta^2).\ea In
the end, the geometric entropy for each fields is found\footnote{
These can be comparable to the entanglement entropy computed in
\cite{NT} (setting $\beta=2\pi\f{L}{V_1}$) \be
S^{scalar}_{ent}=\f{\pi^2N^2}{3\beta},\ \ \ \
S^{vector}_{ent}=\f{\pi^2 N^2}{9\beta},\ \ \ \
S^{fermion}_{ent}=-\f{\pi^2 N^2}{9\beta}.\label{NT} \ee Notice that
these results agree with each other except the gauge field. This
will be due to the subtle issue raised in the paper by \cite{Ka}. If
we literally evaluate the entanglement entropy from the partition
function, we find the result in (\ref{free}) for the vector field.
However, if we eliminate a surface term we get the result in
(\ref{NT}) which is the twice of the real scalar field result.} to
be \be S^{scalar}_{G}=\f{\pi^2N^2}{3\beta},\ \ \ \
S^{vector}_{G}=-\f{2\pi^2 N^2}{9\beta},\ \ \ \
S^{fermion}_G=-\f{\pi^2 N^2}{9\beta}. \label{free}\ee Notice also
that the total sum is vanishing $S^{\CN =4
SYM}_G=S^{scalar}_{G}+S^{vector}_{G}+S^{fermion}_G=0$.

In the high temperature limit, we need to subtract $S_G^P$ from the
above result, where $S_G^P$ is the geometric entropy in the case where
the fermions obey the periodic boundary condition in the $S^1$
direction so that the partition function becomes tr$(-1)^F e^{-\beta
H}$. This is because in the gravity side calculation, we considered
the difference\footnote{This procedure is not necessary in the
analysis of the low temperature region since it become a minor
contribution.} between the result in the AdS black hole and the one
in the thermal AdS.

In the periodic case, we find the total free energy becomes in the
high temperature limit \be -\beta F^{P}_{tot}=\f{\pi^2
N^2(n-1)}{3\beta}, \ee which leads to the entropy \be
S^{P}_{G}=\f{\pi^2 N^2}{3\beta}. \ee We would like to claim the
difference \be \Delta S_G=S^A_{G}-S^P_{G}=-\f{\pi^2 N^2}{3\beta},
\ee should be comparable to the supergravity result (\ref{sugras}),
which differs with each other by the factor $\f{2}{3}$. This is
analogous to the $\f{4}{3}$ factor in a similar ratio of the thermal
entropy \cite{Gub} (a similar ratio in $\CN =1$ conformal field
theories has been worked out recently in \cite{NiTaS}).

In this way, we have shown that the geometric entropy in free $\CN =4$
Yang-Mills qualitatively (or semi-qualitatively) agrees with that in
its holographic dual. Our result provides a strong support that the
geometric entropy is a nice order parameter of Hagedorn/deconfinement
phase transition.

\section{Geometric Entropy in TQFT and 2D YM}
\setcounter{equation}{0} {\hspace{5mm}}
As we have seen, the geometric entropy successfully plays the role
of order parameter in the $\CN =4$ Yang-Mills. Another advantage of
considering this quantity is that we can define the geometric
entropy in any Euclidean field theory, even if the spacetime is not
a direct product of the (Euclidean) time times a space manifold. A
typical such example is the quantum field theory on $S^2$. The
geometric entropy is defined by introducing the deficit angle at two
points on the sphere, e.g. the North and South Pole, in a similar
way we did for $S^3$.

If we consider a two dimensional topological field theory defined on
a Riemann surface $\Sigma_g$, then the partition function $Z_g$
depends only on the genus $g$ and not on the other geometrical
parameter or moduli. We can define the geometric entropy by
introducing a cut on $\Sigma_g$. This leads to the $n$-sheeted
Riemann surface. Then this $n$-sheeted surface has the genus $G=ng$.
The position of the cut is not important as the theory is
topological. In the end, the entropy is defined as follows \be
S_G(g)=-\f{\de}{\de n}\log\left[\f{Z_{ng}}{(Z_{g})^n}\right]\Biggl
|_{n=1}. \ee Especially, in the sphere case $g=0$, we simply find
\be S_G(0)=\log Z_0. \ee A similar result can be found for the
three dimensional topological field theory such as the Chern-Simons
gauge theory on a three sphere. By putting the deficit angle along a
circle, the geometric entropy becomes (see \cite{DFLN}) \be
S_G(S^3)=\log Z(S^3). \ee This is exactly the same as the
topological entanglement entropy introduced in \cite{KP,LW}.

A more interesting example may be the $U(N)$ two dimensional
Yang-Mills theory. It has been shown that the system undergoes a
third order phase transition \cite{Douglas:1993iia} by computing the
partition function exactly using the well-known formula
 \be Z(g,A)=\sum_{R}(\mbox{dim} R)^{2-2g}\
e^{-\f{g^2}{2N}C_2(R)\ti{A}}. \ee Here $\ti{A}$ is the area of the
Riemann manifold; $R$ is a representation of $U(N)$. Below we
measure the area in units of $1/g^2$ i.e. $A=\ti{A}g^2$.

Now we would like to see if the geometric entropy can be regarded as
a order parameter. We concentrate on the genus $0$ case and put a
cut between the North and South pole. Then the geometric entropy is
given by \be S_G(A)=-\f{\de}{\de
n}\log\left[\f{Z(nA)}{(Z(A))^n}\right]\Biggl |_{n=1}=N^2(AF'(A)-F(A)), \ee
where $N^2F(A)=-\log Z(A)$. By employing the analytic expressions of
the free energy $F(A)$ in \cite{Douglas:1993iia}, we can compute the
gap between $S_G$ in the strongly coupled phase $A>A_c$ and the
weakly coupled on $A<A_c$, where $A_c$ is the value of $A$ where the
phase transition occurs. It behaves like \be \Delta S_G(A)\sim N^2
(A-A_c)^2. \ee Therefore, in this example, we can regard
$\f{d^2S_G(A)}{dA^2}$ as an order parameter of the phase transition
in the two dimensional Yang-Mills. In other words, the geometric
entropy is an analogue of the thermodynamical entropy for a quantum
field theory on a general Euclidean manifolds.

\section{Conclusion}
\setcounter{equation}{0} {\hspace{5mm}}
In this paper we introduced a new quantity called the geometric
entropy in quantum field theories, especially focusing on the gauge
theories which often have their holographic duals via AdS/CFT. This
quantity is analogous to the Polyakov loop and indeed we defined it
by a double Wick rotation of another basic quantity known as the
entanglement entropy.

The main claim of this paper is that the geometric entropy can be
used as an order parameter of Hagedorn/deconfinement phase
transitions. We explicitly examined the geometric entropy in both
Yang-Mills theory and its AdS dual and showed that this claim is
indeed true. We also noticed that this quantity plays the role of
order parameter in the two dimensional Yang-Mills theory. It will be
an intriguing future direction to investigate other phase
transitions from the viewpoint of the geometric entropy.

The advantage of considering the geometric entropy is that it is a
universal physical quantity because we can define this quantity in
any quantum field theories even if they are not gauge theories. It
gives much more detailed information than the thermal entropy and
the energy stress tensor do. Therefore it will be very interesting
to understand the holography in more general spacetimes such as the
de-Sitter space by using the geometric entropy as a probe.

\vskip3mm

\noindent {\bf Acknowledgments}

We would like to thank W.~Li for careful reading of this manuscript.
The work of TN is supported in part by JSPS Grant-in-Aid for
Scientific Research No.19$\cdot$3589.
 The work of TT is supported in part by JSPS Grant-in-Aid for
Scientific Research No.18840027 and by JSPS Grant-in-Aid for
Creative Scientific Research No. 19GS0219.

\vskip2mm

\appendix
\section{Computation of the single particle partition functions} \label{aa}
\setcounter{equation}{0} {\hspace{5mm}}
Let's choose the metric of $S^3$ as
\begin{align}
  d\Om^2_{(3)} = d\t^2 + \sin^2\t d\phi^2 + \cos^2\t d\psi^2,
\end{align}
where $0\le \t \le \f{\pi}{2}$ and $0 \le \phi, \psi \le 2\pi$.
This can be embedded in $\mathbb{C}^2$ as
\begin{align}
  z_1 = \sin\t e^{i\phi},\qquad z_2 = \cos\t e^{i\psi}.
\end{align}
When we take the $Z_n$ orbifold of $S^3$, $\psi \simeq \psi +
\f{2\pi}{n}$, the $\mathbb{C}^2$ coordinates become
identified\footnote{A similar calculations in a different orbifold
can be found in \cite{Hi}.} as $z_2 \simeq e^{i\f{2\pi}{n}}z_2$. The
$Z_n$ action acts on the field $\cal O$ \cite{DM}
\begin{align}
  e^{2\pi i\f{a-b}{n}} {\cal O}_{ab} = e^{2\pi i \f{J_2}{n}} {\cal O}_{ab},
\qquad (a,b = 1, \dots, n),
\end{align}
where we denote $SO(4)$ generators as $J_1, J_2$ which act $z_1, z_2$
plane respectively.
Decomposing $SO(4) \simeq SU(2)_L\times SU(2)_R$ and introducing
$m_L \equiv J_1 + J_2$, $m_R \equiv J_1 - J_2$,
the $Z_n$ invariant modes satisfy
\begin{align}
  m_l - m_R = a-b \qquad (mod\ n).
\end{align}
Here we take trivial modes ($a=b$) as invariant states, which
satisfy $m_L-m_R = n\mathbb{Z}$.
The single particle partition function for the scalar field can be represented
as the summation of the invariant state with the representation
$(m_L, m_R)=(j,j)$ and the energy $E=2j+1$
\begin{align}
  z_s(x) 
  &=  6\, \sum_{j=0,1/2.\dots}\sum_{m_L=-j}^{j}\sum_{m_R=-j}^{j}
  x^{2j+1}y^{m_L-m_R}|_{m_L-m_R =n\mathbb{Z},\ y=1}\no
&= 6 \f{x(1+x^n)}{(1-x)^2(1-x^n)}.
\end{align}
Similar calculation leads the single particle partition function
for the vector field
\begin{align}
  z_v(x) &= \sum_{j=0,1/2.\dots}\sum_{m_L=-j-1}^{j+1}\sum_{m_R=-j}^{j}
  x^{2j+2}y^{m_L-m_R}|_{m_L-m_R =n\mathbb{Z},\ y=1} + (m_L \leftrightarrow m_R)\no
  &= \f{2x^2(1+2x^{n-1}-x^n)}{(1-x)^2(1-x^n)}.
\end{align}

For fermions, there are two kinds of the $Z_n$ action
\begin{align}
  g=e^{2\pi i\f{J_2}{n}} \quad or\quad e^{2\pi i\f{n+1}{n}J_2},
\end{align}
but we must take the latter with $k=2l+1$ to require $g^n=1$.
Hence the invariant fermionic states satisfy
\begin{align}\label{Fproj}
  (2n+1) (m_L - m_R) = 0,\qquad (mod\ n),
\end{align}
where we take trivial modes as we did in the bosonic computation.
Since $m_L-m_R$ takes half-integer for the fermion,
(\ref{Fproj}) becomes
\begin{align}
  m_L-m_R = n\left(\mathbb{Z}+\f{1}{2}\right).
\end{align}
The resulting single particle partition function becomes
\begin{align}
  z_f(x) &= 4\left\{\sum_{j=0,1/2.\dots}\sum_{m_L=-j-1/2}^{j+1/2}\sum_{m_R=-j}^{j}
  x^{2j+3/2}y^{m_L-m_R}|_{m_L-m_R =n\left(\mathbb{Z}+\f{1}{2}\right),\ y=1} + (m_L \leftrightarrow m_R)\right\}\no
  &= \f{16x^{1+\f{n}{2}}}{(1-x)^2(1-x^n)}.
\end{align}


\end{document}